\documentclass[12pt]{article}
\usepackage{amssymb,graphicx,epsfig,cite}
\renewcommand\baselinestretch{1.3}

\def\lsim{\raisebox{-.4ex}{\rlap{$\sim$}} \raisebox{.4ex}{$<$}}
\def\gsim{\raisebox{-.4ex}{\rlap{$\sim$}} \raisebox{.4ex}{$>$}}

\begin{document}
\thispagestyle{empty}

 \hfill TIFR-TH/12-27

\begin{center}
{\Large\bf Vacuum Stability Constraints and LHC Searches \\ [2mm] 
for a Model with a Universal Extra Dimension}

{\large\sl Anindya Datta}\,$^{a,1}$
and
{\large\sl Sreerup Raychaudhuri}\,$^{b,2}$
 
{\small
$^a$ Department of Physics, University of Calcutta, \\
92 Acharya Prafulla Chandra Road, Kolkata 700\,009, India \\ [2.5mm]
$^b$ Department of Theoretical Physics, Tata Institute of Fundamental 
Research, \\ 1 Homi Bhabha Road, Mumbai 400005, India.
}
\end{center}

\begin{center} {\Large\bf Abstract} \end{center}
\vspace*{-0.25in}
\begin{quotation}
\noindent If the new boson is lying in the narrow mass range between 
122 - 127 GeV is confirmed to be a Higgs boson, then stability of the 
electroweak vacuum in a minimal model with a universal extra dimension 
(mUED) will require a much lower cutoff for the theory 
than has been envisaged earlier. We show that this low cutoff would lead 
to important changes in much of the mUED phenomenology studied till now. 
In particular, prospects for LHC searches for $n=1$ states are rather 
limited, while resonant $n = 2$ states may go completely undetected. 
Prospects for detection at the ILC and CLIC are less affected.
\end{quotation}

\centerline{\sf Pacs Nos: 14.80.Ec, 14.80.Rt, 13.85.Qk}

\vfill

\centerline{\today}

\bigskip

\hrule
\vspace*{-0.1in}
$^1$ adphys@caluniv.ac.in  \hfill  $^2$ sreerup@theory.tifr.res.in

\newpage

The Higgs sector has long been regarded as an Achilles heel for the 
Standard Model (SM), inasmuch as it calls for the induction of 
elementary scalar fields, for which there was not a shred of 
experimental evidence at the time when the SM was developed. Moreover, 
as soon as we attempt to embed the SM in a higher gauge symmetry, the 
mass of such an elementary scalar becomes unstable, and new physics has 
to be invoked to restore the model to consistency. Despite these obvious 
drawbacks, the simplicity of the Higgs model for electroweak 
symmetry-breaking has been attractive enough for it to be taken with all 
seriousness through several decades of searching for the Higgs boson. 
Now, at long last, it seems that a neutral Higgs boson of mass around 
125~GeV has been discovered at the LHC \cite{Higgs_search}. Precise 
knowledge of the Higgs boson mass means that the last unknown parameter 
in the SM is now known, and since each particle mass in the SM is 
related to a coupling constant, this also means that all the 
interactions in the SM are fully known. More specifically, if $M_H 
\approx 125$~GeV, then the Higgs boson self-coupling $\lambda$ is given, 
at the electroweak symmetry-breaking scale, by $\lambda = M_H^2/2v^2 
\approx 0.129$ .

Assuming that the discovered boson is a Higgs boson, as predicted in the 
SM and most of its extensions, the experimental data tell us that its 
mass must lie in the range $122 - 127$~GeV 
\cite{Higgs_search,Higgs_update}. As it happens, this is a range where 
the Higgs boson leads a somewhat precarious existence, for as we 
increase the energy scale $Q$ above the electroweak scale, there is a 
tendency for the running coupling $\lambda(Q)$ to be driven to smaller 
and smaller values, eventually becoming negative. If this happens, the 
scalar potential becomes unbounded from below, or, in the usual jargon, 
the electroweak vacuum becomes unstable. In the SM, however, this could 
happen at an energy scale in the ballpark of $10^{11}$~GeV 
\cite{Giudice2011} --- which is well above the energy scale at which any 
present or foreseeable terrestrial experiment can be done. Presumably, 
some new physics will appear at a high scale before this point of 
instability is reached, and the vacuum of the new theory will be stable 
--- obviously this would not be directly verifiable by experiment, but 
we can take the corresponding scale as the cutoff when calculating 
quantum corrections in the SM framework.

Since the scale where the new physics must appear is far below the 
Planck scale, a Higgs boson discovery in the range $122 - 127$~GeV could 
indicate the existence of new physics beyond the SM. However, no 
definitive statement can be made in the frame work of the SM unless we 
have a more precise determination of the top quark pole mass 
\cite{Alekhin}. More interestingly, the presence of new particles and 
interactions at or around the electroweak scale in models which go 
beyond the SM can lead to considerable changes in the running of the SM 
parameters, including the scalar self-coupling $\lambda(Q)$. Perhaps the 
most dramatic manifestation of this happens in miminal models with a 
universal extra dimension (mUED), in which a whole set of Kaluza-Klein 
(KK) excitations of the SM particles appears every time we cross a 
threshold $Q \approx n/R$, where $n \in \mathbb{Z}$ and $R$ is the 
radius of compactification \cite{mUED}. In such models, the coupling 
constants run much faster than in most other scenarios, following a 
power-law behaviour \cite{gherghetta,blitz} rather than the slower 
logarithmic running familiar to us in the SM. This power-law running has 
three major consequences, viz.,

\vspace*{-0.2in}
\begin{itemize}

\item The three gauge couplings tend to unify approximately at a scale 
$Q \approx 20R^{-1}$. Since it is usual to take $R^{-1}$ somewhere 
around the electroweak scale, this would bring the grand unification 
(GUT) scale down to the order of 10~TeV, i.e. just beyond the kinematic 
reach of the LHC.

\smallskip

\item If $\lambda \ \gsim \ 0.28$ at the electroweak scale, this 
self-coupling grows with energy scale $Q$ and eventually develops a 
Landau pole at a scale around $45R^{-1}$. Obviously, the cutoff of this 
theory cannot lie beyond this scale. This leads to the well-known {\it 
triviality bound} on the mUED model.

\smallskip

\item If $\lambda \ \lsim \ 0.18$ at the electroweak scale, this 
self-coupling decreases with energy, until it is eventually driven to a 
negative value at an energy scale $Q$ which is generally less than the 
GUT scale in this model. At this point, as explained above, the 
electroweak vacuum becomes unstable. This value of $Q$ may, therefore, 
be referred to as the {\it vacuum stability bound} and we should require 
the theory to be cut off at this scale or lower.

\end{itemize} 

\begin{figure}[h]
\begin{minipage}[h]{3.2in}
In Figure~\ref{fig:cutoff}, we have plotted, as a function of the 
compactification radius $R^{-1}$, the ratio $\Lambda/R^{-1} = \Lambda 
R$. The (blue) shaded region shows the variation in this ratio as the 
mass of the SM Higgs boson is varied from 122~GeV to 127~GeV. Obviously, 
assuming tree-level masses, the number of KK modes with mass $M_n = n/R$ 
which can participate in any process will be given by the nearest 
integer less than the ordinate for a given value of $R^{-1}$, plotted 
along the abscissa. It is clear that this number will only vary between 
3 and 6, and can never reach higher values such as 10 and 20 without 
destabilising the electroweak vacuum. Variation of the top quark mass 
between its experimentally allowed limits results is some minor 
distortion of the curves shown in Figure~\ref{fig:cutoff}, but the 
conclusion remains unchanged.
\end{minipage}
\hskip 10pt
\begin{minipage}[h]{3.0in}
\centerline{\epsfig{file=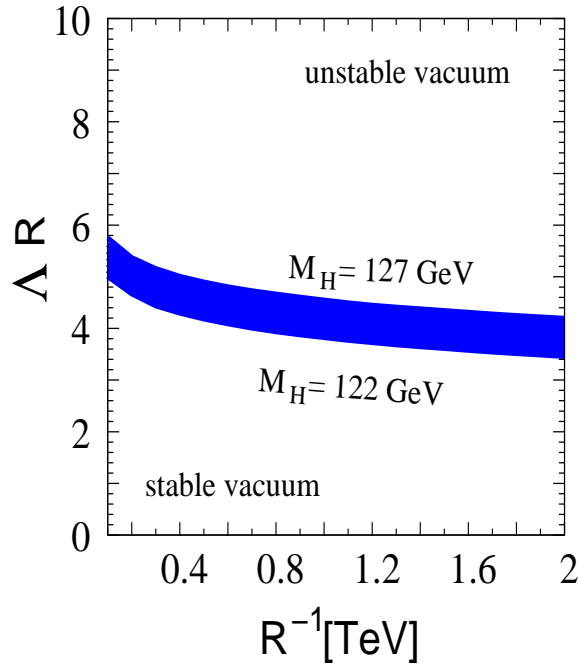,height=3.5in, width = 3.0in}}
\def\baselinestretch{1.2}
\caption{\footnotesize\sl Variation with $R^{-1}$ of the cutoff 
$\Lambda$, presented in terms of the ratio $\Lambda/R^{-1}$, as 
permitted by stability of the electroweak vacuum.}
\def\baselinestretch{1.5}
\label{fig:cutoff}
\end{minipage}
\end{figure}

\vspace*{-0.2in}
The range $0.18 \ \lsim \ \lambda(Q_{\rm ew}) \ \lsim \ 0.28$ is a grey 
area, portions of which can fall under either of the above two cases 
depending on exact value of the top quark Yukawa coupling at the 
electroweak scale, and the accuracy to which the beta functions in the 
theory are evaluated. However, this discussion is now a purely academic 
one, because we know that if the Higgs boson exists, we must have $0.12 
\leq \lambda(Q_{\rm ew}) \leq 0.13$ --- which means that we are 
definitely faced with the scenario described in the third bullet above. 
One therefore, needs to ask the question, what is the vacuum stability 
bound, beyond which the minimal UED model must be cut off, and how many 
KK levels are allowed to contribute to processes generated at the loop 
level? This issue has already been addressed in 
Refs.~\cite{blitz,blennow}, for a somewhat larger range of allowed 
$\lambda(Q_{\rm ew})$ than the above, each of which was consistent with 
the then-current bounds on the mass of the Higgs boson. We revisit the 
bound here, in the light of the narrow window $122 - 127$~GeV in which 
the Higgs boson mas may be presumed to lie. We have used the same 
formulae as in Ref.~\cite{blitz}, with a top quark mass chosen to be 
173.2~GeV. Our results are shown in Figure~\ref{fig:cutoff}.

\bigskip

A similar result also follows from the constraints presented in 
Ref.~\cite{blennow}. What has not been studied in Ref.~\cite{blennow}, 
however, and forms the main thrust of our work, is the serious 
implications such a low cutoff implies for the phenomenology of the mUED 
model in the context of collider searches. In the following discussions, 
we explore these consequences in two different contexts, viz.
\vspace*{-0.2in}
\begin{enumerate}
\item electroweak precision tests; and
\item collider searches for mUED signatures. 
\end{enumerate}  

\vspace*{-0.1in}

Before proceeding further, we pause at this point to recall how the 
masses and couplings of the mUED model are generated. At the tree-level 
the masses $M_n$ of all the KK excitations at the level $n$ ($n \in 
\mathbb{Z}$) are given by
\begin{equation}
M_n^2 = M_0^2 + \frac{n^2}{R^2}
\label{eqn:tree}
\end{equation}
where $M_0$ is the mass of the SM excitation, which corresponds to the 
zero mode $n = 0$ of the KK tower of states. The non-observation of KK 
states at the LEP-2 collider, running at a centre-of-mass energy around 
200~GeV, tells us that the value of $R^{-1}$ is not less than 100~GeV. 
This means that for $n \geq 1$ the KK excitations all have masses in the 
ballpark of $n/R$, i.e. are almost degenerate, with a small splitting, 
due to the zero-mode masses $M_0$. However, this is not the end of the 
story, for each KK mass receives quantum corrections at the loop level, 
i.e. we should re-write Eqn.~(\ref{eqn:tree}) as
\begin{equation}
M_n^2 = M_0^2 + \frac{n^2}{R^2} + \delta M_n^2
\label{eqn:loop}
\end{equation}
where the $\delta M_n^2$ represents the radiative corrections at the 
level of as many loops as we wish to compute. Since the couplings are 
usually of electroweak strength, it is usually sufficient to compute the 
one-loop corrections. However, it is important to note that the loop 
momentum will run all the way up to the cutoff scale $\Lambda$ of the 
theory, i.e. all the KK levels $n$ which can be included within the 
scale $\Lambda$ would contribute to the $\delta M_n^2$ term in 
Eqn.~(\ref{eqn:loop}). The detailed formulae for these corrections 
$\delta M_n^2$ may be found in Ref.~\cite{Matchev}, and have been used 
for our numerical estimates. These formulae are also implemented in a 
software package \cite{Asesh} to be run in conjunction with the Monte 
Carlo event generator {\sc CalcHEP} \cite{calchep}, and we have verified 
that our results are in good agreement with this package.

\begin{figure}[ht]
\centerline{\epsfig{file=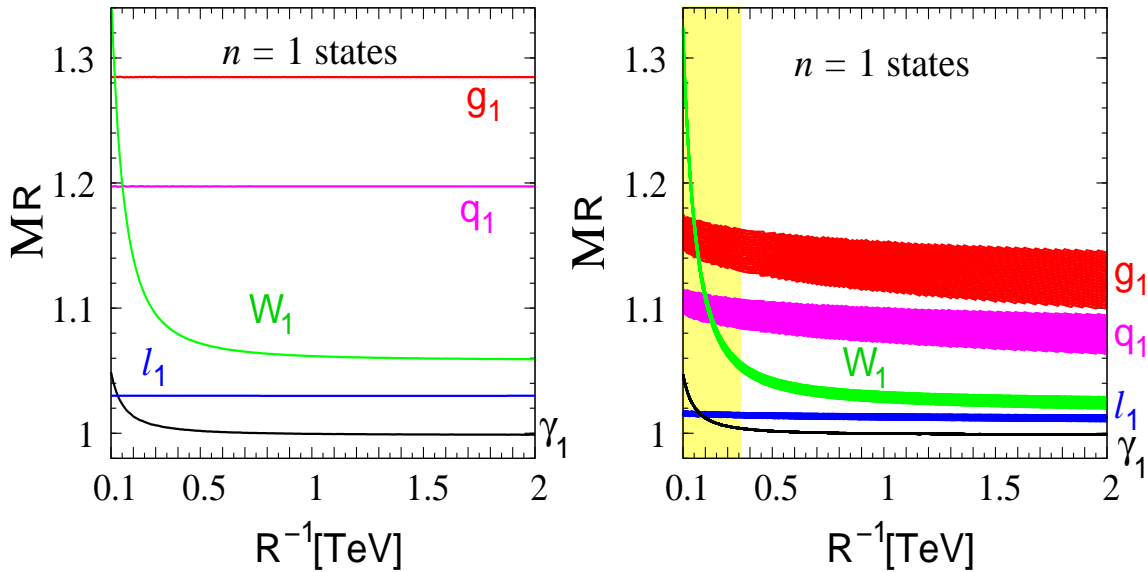,height=3.0in,width=6.0in}}
\def\baselinestretch{1.2}
\caption{\footnotesize\sl Variation with $R^{-1}$ of the masses of some 
of the $n = 1$ KK excitations. The lines/bands coloured red, pink, 
green, blue and black correspond, respectively, to the masses of the 
$g_1$, $q_1$, $W_1$, $e_1$ and $\gamma_1$. Each mass is presented as a 
multiple of the common scale $R^{-1}$. The right panel shows the 
spectrum with summation over the allowed number of KK levels, while the 
left panel shows the spectrum with summation over 20 KK levels. The 
yellow-shaded region in the right panel indicates the region $R^{-1} 
\leq 260$~GeV, which is more-or-less ruled out by LEP data on the 
oblique parameters (see below).}
\def\baselinestretch{1.5}
\label{fig:spectrum}
\end{figure}

Generally, the phenomenological results available in the literature use 
a cutoff $\Lambda \approx 20R^{-1}$, which is consistent with the GUT 
scale in a mUED scenario. Now, however, we have seen that the cutoff 
must be much smaller, with $\Lambda \ \lsim \ 4R^{-1}$ instead. Thus, 
when only three or four KK levels contribute to the self-energy 
corrections in place of twenty, it is natural to assume that these 
radiative corrections will be significantly smaller than those predicted 
earlier. Since it is primarily the $\delta M_n^2$ which lift the 
degeneracy between different states at the same $n$ level, one may 
expect the states to be rather more degenerate than had been thought 
earlier. This is illustrated in Figure~\ref{fig:spectrum}, where we show 
the variation of the $n = 1$ masses, as a function of the ratio 
$M_1/R^{-1} = M_1 R$, when $R^{-1}$ varies from 100~GeV to 2.0~TeV.

The different coloured bands in Figure~\ref{fig:spectrum} correspond to 
variation of some of the important masses in the $n = 1$ KK spectrum of 
the mUED model, viz., the $n = 1$ excitations of the gluon ($g_1$), the 
light quarks ($q_1$), the $W$-boson ($W_1$), the electron ($e_1$) and 
the photon $\gamma_1$. Of these, the $\gamma_1$, whose mass remains 
more-or-less exactly at $R^{-1}$ is clearly the LKP. The thickness of 
the bands in the right panel corresponds, as in Figure~\ref{fig:cutoff}, 
to variation of $M_H$ over the allowed range, with the lower~(upper) 
edge corresponding to $M_H = 122 \ (127)$~GeV. What strikes us 
immediately about the spectrum is that at the lower end, the overall 
splitting between the lightest KK particle (LKP), viz., the $\gamma_1$ 
and the heaviest $n = 1$ excitation, viz., the $g_1$, is never more than 
about 17\%. By contrast, if we sum the loop corrections over 20 KK 
levels, we would predict a maximum splitting around 30\%, i.e roughly 
double of what we see in Figure~\ref{fig:spectrum}. To take a concrete 
example, if $R^{-1} = 500$~GeV, the mass splitting is never more than 
about 80~GeV, whereas if we had summed over 20 levels, we would have 
predicted a splitting close to 150~GeV. It is interesting that the only 
specific input in all of this is the newly-measured mass of the Higgs 
boson -- the rest follows inexorably from the vacuum stability argument. 
Such compression of the spectrum will happen in all the KK levels, 
though we are most affected by the compression in the first level.

\bigskip

\begin{figure}[htb]
\centerline{\epsfig{file=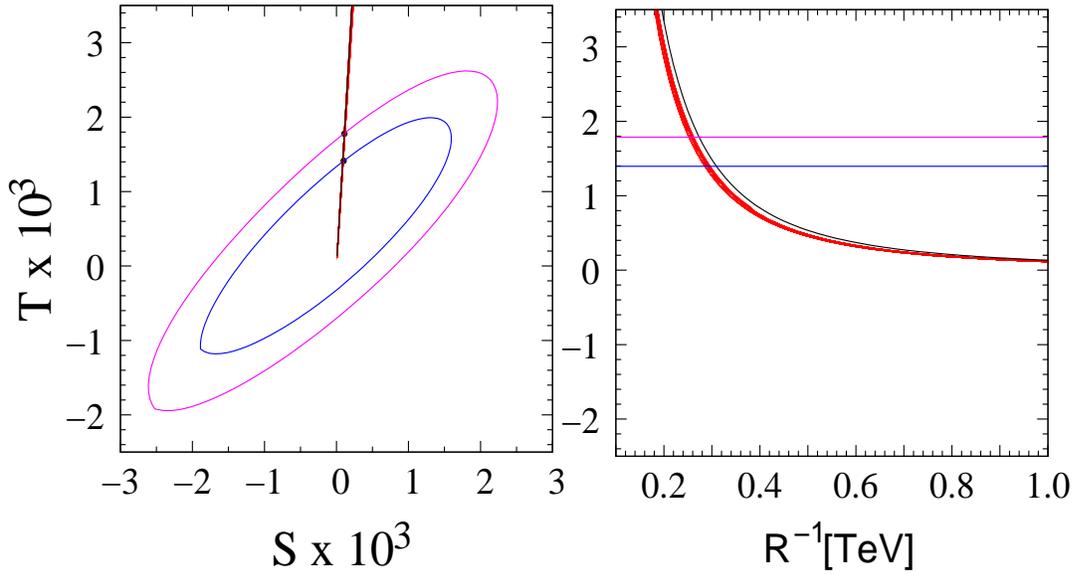,height=3.0in,width=5.6in}}
\def\baselinestretch{1.2}
\caption{\footnotesize\sl Oblique parameters in the mUED model. Both 
panels show the mUED contributions in red~(black) depending on whether 
the summation over KK levels is carried out to realistic~(twenty) 
levels. The ordinate is $T \times 10^3$ in both panels. The blue~(pink) 
ellipses on the left panel correspond to the LEP-2 constraints at 
95~(90)~C.L. and the blue~(pink) lines on the right panel indicate the 
values of ordinate for which these are crossed by the mUED 
contributions, i.e. all values of ordinate lower than these lines are 
permitted by the electroweak data at 95~(90)~C.L..}
\def\baselinestretch{1.5}
\label{fig:oblique}
\end{figure}

Let us now take up the consequences of this effect in the specific 
phenomenological contexts mentioned above. Obviously, the oldest of 
these constraints come from the electroweak precision tests -- more 
specifically, from the mUED model contributions to the oblique 
parameters $S$, $T$ and $U$ \cite{peskin}. Though these have been 
calculated by many authors \cite{oblique}, we use the simple and 
accurate formulae presented in Ref.~\cite{STU}. Once again, we set the 
mass of the top quark to 173.2~GeV in our calculation. This is, in fact, 
an important parameter in the calculation of the oblique parameters, but 
we have checked that variation over the allowed experimental range 
produces very small effects so far as the constraints on the mUED model 
are concerned. It is, therefore, convenient to keep its value fixed and 
vary other important parameters, such as the Higgs boson mass $M_H$ and 
the compactification radius $R$.

In Figure~\ref{fig:oblique} we plot the deviations $S$ and $T$ from the 
SM values, so that the SM corresponds to the origin of this graph. A 
glance at the left panel of Figure~\ref{fig:oblique} informs us that the 
mUED contribution to the $S$ parameter is so small that the constraint 
on this model arises essentially from the $T$-parameter, or, to use its 
older equivalent, the $\rho$-parameter. The $S$ parameter contributes 
nevertheless, to the constraint through its correlation with the 
$T$-parameter, which is obvious from the fact that the allowed regions 
form ellipses rather than rectangles in the plots. Thus, the limits on 
the mUED model may be read off from the plot by considering the 
intersection of the mUED plot with the 90\% and 95\% C.L. ellipses. If 
we do this, we can demand that $T$ will be less than 1.788 (1.396) at 90 
(95)\% level. Now, the black plots indicate the mUED contributions when 
we sum over 20 levels, while the red plots indicate the same results 
with a realistic cutoff. The thickness of the red curves indicates the 
result of varying $m_H$ over the range $122 - 127$~GeV.

On the right panel of Figure~\ref{fig:oblique}, we have plotted the mUED 
contribution to the $T$ parameter as a function of $R^{-1}$, with red 
(black) curves indicating sums over realistic (twenty) levels. The 
difference is rather small, and hence, the imposition of vacuum 
stability constraints on the mUED model results in a slight decrease in 
the constraint on $R^{-1}$. The exact results are given in 
Table~\ref{tab:STU} below.

\begin{table}[h]
\begin{center}
\begin{tabular}{ccc}
\hline
levels & 95\%C.L.  & 99\%C.L. \\
\hline\hline
$3 - 4$ & 285 (296) & 249 (261) \\
20      & 310        & 273        \\
\hline
\end{tabular}
\def\baselinestretch{1.2}
\caption{\footnotesize\sl Lower limits (in GeV) arising from electroweak 
precision tests, on the compactification scale $R^{-1}$. The first line 
shows the realistic bounds, with the range corresponding to $M_H = 122 
(127)$ GeV respectively, while the second line corresponds to summing 
over 20 levels --- which is unrealistic, but consistent with earlier 
practice.}
\def\baselinestretch{1.5}
\label{tab:STU}
\end{center}
\end{table}

\vspace*{-0.2in}
The numbers presented in Table~\ref{tab:STU} make it clear that the 
compactification scale in the mUED model must lie above the electroweak 
scale, viz., 246~GeV. However, the difference induced by the vacuum 
stability constraint are rather modest, being at a level less than 10\% 
irrespective of the mass of the Higgs boson. On the whole, therefore, we 
may say that electroweak precision constraints on the mUED model 
essentially stand, with a slight relaxation from the earlier results.

The compression of the spectrum has much more dramatic effects when we 
consider collider searches for the KK excitations in the mUED model. In 
this context, it is first worth noting that the constraints from 
precision electroweak data indicate that direct searches at both the 
LEP-2 and Tevatron colliders would fail, as pair-production of even the 
$n = 1$ KK excitations is beyond the kinematic limit of these machines. 
It must be to the LHC, running in the range of several TeV, that we turn 
for these direct searches. At this point, a quick review of mUED search 
strategies at the LHC is needed before we can assess the impact of the 
compression effect on these searches.

Production modes for KK excitations at the LHC can be divided into two 
classes, both requiring roughly the same amount of energy, viz.,
\vspace*{-0.2in}
\begin{itemize}

\item pair-production of $n = 1$ modes, which requires energy in the 
ballpark of $2 \times R^{_1}$, since the masses of $n = 1$ modes is in 
the ballpark of $R^{-1}$;

\item resonance production of $n = 2$ modes, which requires energy in 
the ballpark of $1 \times 2R^{_1}$, since the masses of the $n = 2$ 
modes is in the ballpark of $2R^{-1}$.

\end{itemize}

mUED searches based on the first of these, namely pair-production of $n 
= 1$ states, are very reminiscent of searches for supersymmetry, 
resulting in the appellation 'bosonic supersymmetry' for the mUED model. 
The principal modes for such pair production involve the 
strongly-interacting $n=1$ states and are
\begin{eqnarray}
pp \to \left\{ \begin{array}{l} g_1 + g_1 \\ 
                                q_1 + \bar{q}_1 \\ 
                                g_1 + q_1 (\bar{q}_1) \end{array} \right.
\label{eqn:strong}
\end{eqnarray}
where $q$ stands for any quark flavour, either singlet or doublet, with 
the top quark included if it is kinematically possible. The $g_1$ will 
always decay as
\begin{eqnarray}
g_1 & \to \left\{ \begin{array}{l} q + \bar{q}_1 \\ 
                                  q_1 + \bar{q} \end{array} \right.
\end{eqnarray}
with a hadronic jet arising from the $q$ or $\bar{q}$, as the case may 
be. Obviously, since the gluon coupling is vectorlike, equal numbers of 
doublet $q_{1L}$ and singlet $q_{1R}$ states will be produced. Decays of 
these $q_1$ states involve more channels, viz.,
\begin{eqnarray}
q_{1L} & \to \left\{ \begin{array}{l} 
q + \gamma_1 \\ 
q + Z_1^0  \to q + \ell^\pm + \ell^\mp_1 \to q + \ell^+\ell^- + \gamma_1 \\                                 
q' + W_1^\pm \\
     \hspace*{0.2in}   \begin{array}{rll}  
\hookrightarrow & W^\pm + \gamma_1 & \to \ell^\pm + \nu(\bar{\nu}) + \gamma_1 \\
 &  & \hookrightarrow q\bar{q} + \gamma_1 \\
\hookrightarrow & \nu(\bar{\nu}) + \ell^\pm_1 & \to \nu(\bar{\nu}) + \ell^\pm + \gamma_1 \\
\hookrightarrow & \ell^\pm + \nu_1(\bar{\nu}_1) & \to \nu(\bar{\nu}) + \ell^\pm + \gamma_1
                       \end{array} 
                   \end{array}  \right.
\label{eqn:q1Ldecay}
\end{eqnarray}
and
\begin{eqnarray}
q_{1R} & \to \left\{ \begin{array}{l} 
q + \gamma_1 \\ 
q + Z_1^0  \to q + \ell^\pm + \ell^\mp_1 \to q + \ell^+\ell^- + \gamma_1  
                   \end{array}  \right.
\label{eqn:q1Rdecay}
\end{eqnarray}
with the final states in Eqn.~\ref{eqn:q1Ldecay} being, from top to 
bottom, hadronic jet plus missing $E_T$ (MET), hadronic jet plus 
dilepton plus MET, single lepton plus MET, pair of jets plus MET, single 
slepton plus MET, single lepton plus MET; while the final states in 
Eqn.~\ref{eqn:q1Rdecay} are, from top to bottom, hadronic jet plus 
missing $E_T$ (MET) and hadronic jet plus dilepton plus MET. Here we 
must remember that the number of jets is only a rough estimate, since 
jets can always split and/or merge during the fragmentation process.

When we consider all the possible cascade decays of the pair of $q_1, 
g_1$ states produced, it is obvious that the final states will always 
involve ($a$) large missing transverse energy (MET), and ($b$) an 
indeterminate but limited number of hard leptons and jets --- where the 
word `hard' is used not so much in the usual sense of large transverse 
momentum ($p_T$) but more to distinguish these from `soft' leptons and 
jets produced during fragmentation. The reason for this last caveat is 
that the transverse momentum carried by these arises almost wholly from 
the mass difference between the parent and daughter states, which, in a 
compressed mUED mass spectrum, is not very large. The question then 
arises as to whether these leptons and hadronic jets can clear the 
minimum $p_T$ cuts employed in mUED searches. To study, this, it is 
clear that the relevant mass differences, with the corresponding states, 
are
\vspace*{-0.2in}
\begin{enumerate}
\item jet: $M(g_1) - M(q_1)$, or, $M(q_1) - M(\gamma_1)$, or, $M_(q_1) - M(W_1)/M(Z_1)$;
\item lepton: $\frac{1}{2}\left[ M(Z_1) - M(\ell_1) \right]$, or, 
$\frac{1}{2}\left[ M(W_1) - M(\gamma_1) \right]$, or, 
$M(\ell_1) - M(\gamma_1)$.
\end{enumerate}
\vspace*{-0.2in}
We must note that these mass differences form approximate upper bounds 
on the transverse momentum $p_T$; for most events the actual $p_T$ value 
is lower. We also note in passing that though we will present mass 
differences for the choice $q_1 = u_{1L}$, there will be very little 
change if any of the others is chosen.

\begin{figure}[htb]
\centerline{\epsfig{file=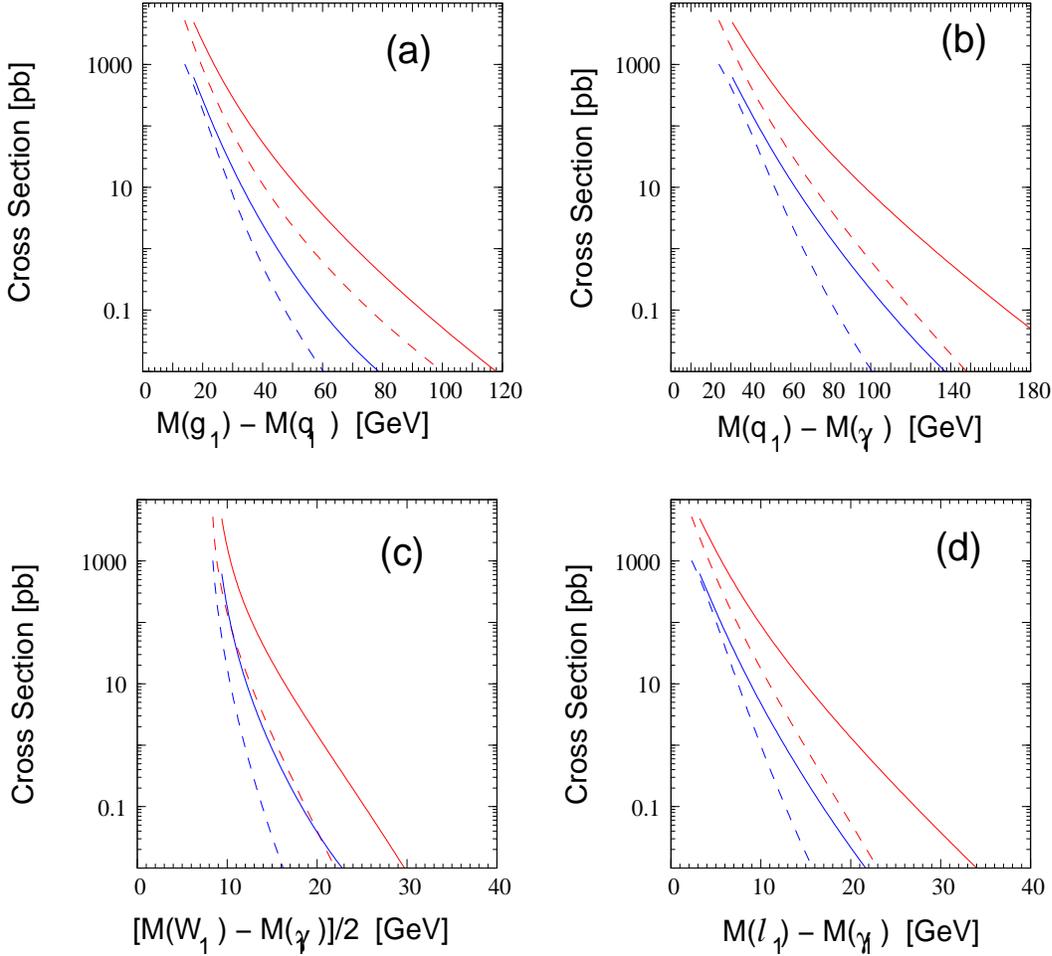,height=5.0in,width=5.5in}}
\def\baselinestretch{1.2}
\caption{\footnotesize\sl Cross-sections for $g_1$ and $q_1$ pair 
production (see Eqn.~(\ref{eqn:strong})) at the LHC, versus 
various mass differences as labelled corresponding to the maximum 
energies of ($a,b$) jets, ($c$) jets and leptons, and ($d$) leptons, 
arising in the cascade decays of $n = 1$ KK resonances. Solid (broken) 
lines correspond to $M_H~=~127~(122)$~GeV, while the colours red(blue)
correspond to $\sqrt{s}~=~14~(7)$~TeV respectively. }
\def\baselinestretch{1.5}
\label{fig:lhc}
\end{figure}

One of the most important results of having a compressed spectrum is 
that for a given value of $R^{-1}$, the masses of the $q_1$ and $g_1$ 
states are lighter than before, and hence, these would be produced more 
copiously at the LHC. To illustrate this, as well as the upper bounds on 
jet and lepton $p_T$, we have plotted, in Figure~\ref{fig:lhc}, the 
total production cross-sections at the 14(7)~TeV LHC for the processes 
listed in Eqn.~\ref{eqn:strong} versus the mass differences \\
\hspace*{0.2in} (a) \ $M(g_1) - M(q_1)$; \\
\hspace*{0.2in} (b) \ $M(q_1) - M(\gamma_1)$; \\
\hspace*{0.2in} (c) \ $\frac{1}{2}\left[ M(W_1) - M(\gamma_1) \right]$; \\
\hspace*{0.2in} (d) \ $M(\ell_1) - M(\gamma_1)$. \\
In every plot, the solid~(broken) lines indicate a Higgs boson mass of 
127~(122)~GeV, while the colour red(blue) corresponds to 14 (7)~TeV. The 
value of $R^{-1}$ varies from 300~GeV to several TeV as we proceed 
downwards along each line. Cross-sections obviously tend to decrease as 
the Higgs boson mass increases, and this is accentuated at larger values 
of $R^{-1}$.

A striking feature of the plots shown in Figure~\ref{fig:lhc} is the 
fact that the raw cross-sections for production of $n = 1$ particles can 
be very large -- several nanobarns for low values of $R^{-1}$, but fall 
rapidly as $R^{-1}$ increases. This is not surprising, for after all, it 
is known that the $t\bar{t}$ production cross-section is close to a 
nanobarn, and the heavy $q_1$ states are, after all, very similar to top 
quarks. For the same reason, however, $t\bar{t}$ production could be an 
irreducible background to mUED signals, since the top quark and its 
anti-quark are known to decay to leptons, hadronic jets and large MET. 
In, fact, to get rid of the $t\bar{t}$ and other SM backgrounds, 
including the enormous QCD background, it is usual to impose some 
minimum cuts on the $p_T$ of these final states. This, we shall see, 
will deeply affect the kind of searches we are now considering. For, as 
the plots in Figure~\ref{fig:lhc} show, the $p_T$ of the final states 
leptons and jets is severely limited by the small mass differences in 
the compressed spectrum. As a result, $p_T$ cuts imposed to remove SM 
backgrounds will also remove substantial portions of the mUED signals. 
In fact, the low cross-sections for $\sqrt{s} = 7$~TeV, illustrated in 
Figure~\ref{fig:lhc}, lead us to believe that we must look to the 14~TeV 
option to see any viable signal.

Let us consider the mUED signals in some more detail. If we go by the 
analogy of supersymmetry, the most promising final state should be 
multiple jets and large missing transverse energy (MET). Typical $p_T$ 
triggers for multi-jet plus MET signals in supersymmetric models 
generally lie at the level of 50 to 100 GeV, and it is clear from 
Figure~\ref{fig:lhc}~($a$) and ($b$) that the mUED signal events would 
be able to clear such trigger requirements only for larger values of 
$R^{-1}$, where the cross-sections are rather small. However, if the 
trigger requirements are relaxed in a low-$p_T$ search, the situation 
does not improve substantially, for then the SM backgrounds become 
intractable. To make matters concrete, we carry out a Monte Carlo 
simulation of the signal with multiple jets and MET at $\sqrt{s} = 
14$~TeV, using the well-known Monte Carlo event generator {\sc Pythia} 
\cite{pythia}, for $R^{-1} = 500$~GeV, with weak and strong kinematic 
criteria as follows: \\
\hspace*{0.5in} C0: only lepton veto; \\
\smallskip
\hspace*{0.5in} C1: lepton veto, $p_T^J \geq 20$~GeV and $\not{\!\!\!E}_T \geq 30$~GeV; \\ 
\smallskip
\hspace*{0.5in} C2: lepton veto, $p_T^J \geq 50$~GeV, $\not{\!\!\!E}_T \geq 100$~GeV. \\
\smallskip
The 'lepton veto' requires the absence of isolated leptons with 
transverse momentum greater than 10~GeV. Our results are given in 
Table~\ref{tab:CShad} below.
    
\begin{table}[h]
\begin{center}
\begin{tabular}{crrrrr}
\hline
 cuts   & model  & $\leq 2$ jets + $\not{\!\!E}_T$ & 3 jets + $\not{\!\!E}_T$ & 4  jets + $\not{\!\!E}_T$ &  $> 4$ jets + $\not{\!\!E}_T$  \\
\hline\hline
C0 & mUED & 116.9 (97.4) & 66.6 (63.4) & 44.2 (41.5) & 37.9 (43.8) \\ \hline
C1 & mUED & 86.4 (74.8)  & 55.0 (53.3) & 39.1 (36.9) & 35.6 (40.8) \\  
     & SM ($t\bar{t})$   &   2.9    &  7.4    & 12.3  & 23.3    \\ [-1mm]
     & SM (QCD)          &  1432.4  &  1386.3  & 850.1 & 612.5   \\  
\hline
C2 & mUED & 4.56 (3.12) & 1.08 (1.4) & 0.58 (0.67) & 0.16 (0.31) \\       
\hline
\end{tabular}
\def\baselinestretch{1.2}
\caption{\footnotesize\sl The mUED signal in pb corresponding to $R^{-1} 
= 500$~GeV and $M_H = 122~(127)$~GeV and SM backgrounds in the 
multi-jets + MET channels at the 14~TeV LHC, showing the effect of 
selection criteria defined by {\rm (C0)} only lepton veto; {\rm (C1)} 
lepton veto, $p_T^J \geq 20$~GeV, $\not{\!\!\!E}_T \geq 30$~GeV; {\rm 
(C2)} lepton veto, $p_T^J \geq 50$~GeV, $\not{\!\!\!E}_T \geq 
100$~GeV. All detection efficiencies are assumed to be unity.}
\def\baselinestretch{1.5}
\label{tab:CShad}
\end{center}
\end{table}

\vspace*{-0.2in}
It is clear from Table~\ref{tab:CShad} that, irrespective of the exact 
mass of the Higgs boson, the mUED signal is at the level of a couple of 
hundred picobarns. This may appear sizeable, but it is not really so 
when we consider the SM backgrounds, of which we have presented just 
two, viz. $t\bar{t}$ production, and QCD production of jets and MET.  
Merely taking these two signals adds up to a few {\it nanobarns}, 
against which there is no hope of discerning the mUED signals. We can, 
of course, reduce the SM background by imposing more stringent kinematic 
cuts on the jets as well as on the MET. However, these will prove more 
deadly for the tenuous signal than for the background. For example, 
imposing $p_T^J \geq 50$~GeV and requiring MET greater than 100 GeV 
effectively reduces the signal by one, or even two, orders of magnitude, 
thereby offsetting any advantage to be gained by imposing stricter 
kinematic criteria on the background. One may safely conclude, 
therefore, that even though $n = 1$ KK states will be copiously produced 
at the LHC, it will be very difficult to isolate them from the 
background, in the multi-jets plus MET channel.

Recently, a combination of event shape variables has been used rather 
effectively \cite{monalisa} to reduce the SM backgrounds to the signal 
with multiple jets and MET. This has proved very successful for searches 
for supersymmetry, where the final state jets are hard and there is 
considerable MET. An allied study \cite{sujoy} of mUED signals using the 
same technique claims that this can be effectively used to search for 
mUED in the jets plus MET channel. However, we note that the success of 
this technique depends rather crucially on the `hardness" of the jets, 
and Ref.~\cite{sujoy} achieves this by taking $\Lambda R = 10$ and 40. 
We suspect that if the same analyses were to be re-done using the 
correct values of $\Lambda R$ as presented in this article, the result 
would prove as disappointing as that for conventional searches described 
above.

We now turn to leptonic signals for mUED. Naively, a glance at 
Figure~\ref{fig:lhc} would make matters seem very gloomy for such 
signals, since the $p_T$ of the leptons, whose upper bounds correspond 
to the abscissa in Figure~\ref{fig:lhc} ($c$) and ($d$), rises above the 
trigger level of 20~GeV only when the cross-section falls to very low 
values. However, if the trigger level is reduced to, say, 10~GeV, the 
mUED signal will be less affected, but will now have to compete with SM 
backgrounds involving soft leptons arising from electroweak sources as 
well as hadronic decays inside jets. It is difficult to guess what will 
happen without performing a detailed study, and hence, we have again 
performed a simulation of the multi-lepton signals at $\sqrt{s} = 
14$~TeV, using {\sc Pythia}, for $R^{-1} = 500$~GeV, and imposing 
various kinematic criteria: \\
\smallskip
\hspace*{0.5in} C1: $p_T^\ell \geq 10$~GeV, $\not{\!\!\!E}_T \geq 20$~GeV; \\
\smallskip
\hspace*{0.5in} C2: $p_T^\ell \geq 20$~GeV, $\not{\!\!\!E}_T \geq 30$~GeV;  \\
\smallskip
\hspace*{0.5in} C3: $p_T^\ell \geq 20$~GeV, $\not{\!\!\!E}_T \geq 50$~GeV; \\
\smallskip
\hspace*{0.5in} C4: 10~GeV~$\leq p_T^\ell < 25$~GeV, 10~GeV~$\leq~\not{\!\!\!E}_T < 25$~GeV. \\
Our results are illustrated in Table~\ref{tab:CSlep}.

A close study of Table~\ref{tab:CSlep} reveals the following features. 
Focussing on the mUED signal, it is clear that one may expect reasonably 
large cross-sections for signals with leptons, jets and MET, with lepton 
multiplicities of 1, 2 and 3, but a somewhat diminished probability of 
producing 4 leptons. This is easy to motivate from the decay chains 
listed in Eqns.~\ref{eqn:q1Ldecay} and \ref{eqn:q1Rdecay}. However, the 
application of even so gentle a kinematic selection criterion as 
C1:~$p_T^\ell \geq 10$~GeV, $\not{\!\!\!E}_T \geq 20$~GeV leads to a 
considerable diminution of the signal in all channels except the one 
with a single lepton where the numbers are obviously not competitive 
with the SM backgrounds shown immediately below, which arise mainly from 
the production and decay of $W,Z$ states, with or without associated 
jets, from $t\bar{t}$ production, and from QCD jets, which generate a 
hard isolated lepton only once in a few millions, but nevertheless form 
a substantial background because of the enormous cross-section for QCD 
processes. The QCD background is negligible for the case of two or more 
leptons, but there is a large background to dileptons from electroweak 
processes including the $t\bar{t}$ decays.

\begin{table}[h]
\begin{center}
\begin{tabular}{cr|rr|rr}
\hline
& &  $1\ell$ + jets + $\not{\!\!E}_T$ & $2\ell$  + jets + $\not{\!\!E}_T$ &  $3\ell$ + jets + $\not{\!\!E}_T$ &  $4\ell$  + jets + $\not{\!\!E}_T$ \\
 cuts & model & $\sigma$\,(pb) & $\sigma$\,(pb) & $\sigma$\,(fb) & $\sigma$\,(fb) \\
\hline\hline
C1 & mUED & 76.6 (81.8) & 10.39 (11.9) & 725.3 (1\,002.0) & 54.0 (67.2) \\ [-1mm] 
   & SM (EW, $t\bar{t}$)   & 62\,758.0 & 5\,041.9  &  90.8 & 3.5 \\ [-1mm]
   & SM (QCD)              & 30\,420.0 & 0.0091 & -- & -- \\ \hline
C2 & mUED & 11.56 (14.3) & 0.43 (0.7) & 14.1 (7.0) & 1.1 (0.9) \\ [-1mm]  
   & SM (EW, $t\bar{t}$)  &  52\,612.7 & 3\,884.2  &  77.2 & 2.3 \\ \hline
C3 & mUED & 9.47 (11. 2) & 0.31 (0.5) & 14.0 (3.5) & 7.1 (0.4) \\ [-1mm] 
   & SM (EW, $t\bar{t}$)  & 52\,612.7 & 3\,884.2  &  77.2 & 2.3 \\ \hline 
C4 & mUED & 10.35 (9.5) & 1.27 (1.4) & 54.5 (91.4) & 2.0 (11.3) \\ [-1mm]  
   & SM (EW, $t\bar{t}$)  & 16\,434.1 &   892.1  &  0.5 & 0.1 \\ 
\hline
\end{tabular}
\def\baselinestretch{1.2}
\caption{\footnotesize\sl The mUED signal corresponding to $R^{-1} 
= 500$~GeV and $M_H = 122~(127)$~GeV and SM backgrounds in the 
multi-lepton + jets + MET channels at the 14~TeV LHC, showing the effect 
of selection criteria defined by 
{\rm (C1)} $p_T^\ell \geq 10$~GeV, $\not{\!\!\!E}_T \geq 20$~GeV; 
{\rm (C2)} $p_T^\ell \geq 20$~GeV, $\not{\!\!\!E}_T \geq 30$~GeV; 
{\rm (C3)} $p_T^\ell \geq 20$~GeV, $\not{\!\!\!E}_T \geq 50$~GeV;
{\rm (C4)} 10~GeV~$\leq p_T^\ell < 25$~GeV, 10~GeV~$\leq~\not{\!\!\!E}_T < 25$~GeV.
All detection efficiencies are assumed to be unity.}
\def\baselinestretch{1.5}
\label{tab:CSlep}
\end{center}
\end{table}

\vspace*{-0.2in}
Very different from the case of one or two leptons are the two columns 
on the right of Table~\ref{tab:CSlep}. Here, the trilepton and 
quadrilepton signals stand out clearly over the backgrounds, when the 
milder selection criteria C1 and C4 are taken (the converse is the case 
when we take more conventional criteria as in C3 and C4). For the 
trilepton signal, the signal will be substantial even in the early runs 
of the 14~TeV LHC upgrade; for the quadrilepton signal, the signal is 
more modest, and will require perhaps a year or two to show up in enough 
numbers to come to any definite conclusion.  .

Searches for the mUED model at the LHC will, therefore, clearly have to 
wait a few years before any conclusion can be reached. Noting that all 
our results are pertinent to a comparatively low $R^{-1} = 500$~GeV, a 
negative result will only push up the lower bound on $R^{-1}$ to a 
higher value, perhaps eventually in the ballpark of 1.5~TeV. On the 
other hand a positive signal will hardly be distinguishable from those 
of a supersymmetric model with a similarly compressed spectrum. It may 
be mentioned in passing that it has been claimed \cite{inverse} that the 
multiplicity of accompanying jets can be used as a discriminant between 
the mUED and supersymmetric models. Such claims are, however, based on 
studies with a less compressed spectrum and much more stringent 
kinematic cuts on the leptons, accompanying jets, and MET. Their 
applicability to a compressed mass spectrum as considered in this work 
is, therefore, an open question.

Another way to easily distinguish mUED signals from supersymmetric ones 
involves looking for the $n = 2$ KK excitations. Here, however, the 
situation is really gloomy. Once again, the spectrum will be compressed 
-- at the same relative level as that for $n = 1$ states, though the 
splitting would be roughly double that for $n = 1$ states. Referring to 
Figure~\ref{fig:spectrum}, we see that for $R^{-1} = 1$~TeV, we can 
expect a $\gamma_2$ of mass around 2.0~TeV, a $W_2$ and a $Z_2$ of mass 
around 2.08~TeV and a $g_2$ of mass around $2.22 - 2.26$~GeV. These 
states can be produced singly as resonances in quark or gluon fusion, 
and can be detected through their subsequent decay into $\ell^+\ell^-$ 
pairs \cite{Kong} or $t\bar{t}$ pairs \cite{Biplob}.

Compression of the mass-spectrum does not really matter for the single 
production of resonances, but in such studies, what really matters is 
the coupling of the resonance to the partons, which is achieved through 
KK number-violating operators generated at the one-loop level. 
Obviously, such couplings, like the mass-splitting, will be seriously 
affected by the number of KK levels over which the loop momentum is 
summed. An estimate of this effect may be obtained from Figure~2 of 
Ref.~\cite{Biplob}, where the sum over 5 KK levels may be taken to 
closely approximate the effect of the sum over 3--4 levels induced by 
the vacuum stability condition. At a rough estimate, the result of the 
lower number of KK levels summed leads to a reduction of the couplings 
of the $g_1$ to partons by a factor around 4 to 5, i.e. a reduction in 
the resonant cross-section by a factor around 20. If we apply this to 
the results shown (for $\Lambda R = 20$) in Figure~6 of 
Ref.~\cite{Biplob}, the nice resonances shown therein shrink to levels 
which are smaller even than the $1\sigma$ fluctuation in the $t\bar{t}$ 
background from the SM. As the relevant plot already assumes a 
luminosity of 100~fb$^{-1}$ at 14~TeV, it is more-or-less obvious that 
$g_2$ resonances, like the $n = 1$ cascade products, will be lost 
against the SM backgrounds. A similar argument may be applied to 
$\gamma_2$ and $Z_2$ resonances, decaying to $e^+e^-$ and $\mu^+\mu^-$ 
pairs, illustrated in Figure~9 of Ref.~\cite{Kong}. In this case, the 
reduction in the coupling is by a factor around 10 rather than 20, but 
this will still leave barely one or two signal events to be discovered 
in 100~fb$^{-1}$ of data, against a background in the ballpark of $10 - 
20$ events. Thus, the search for $n = 2$ states seems to be another issue 
without hope.

It is quite clear, therefore, that if a Higgs boson is discovered in the 
now-relevant mass range of $122 - 127$~GeV, this will immediately render 
sterile all prospects of finding mUED signals at the LHC, except in the 
trilepton and quadrilepton channels --- where they will be 
indistinguishable from a class of models with supersymmetry. One can, 
then ask the legitimate question whether it is better worth looking for 
such signals at the proposed International Linear Collider (ILC), where 
$e^+e^-$ pairs will collide at energies of 500~GeV or 1~TeV. Indeed 
$e^+e^-$ colliders are known to be much cleaner as regards SM 
backgrounds, and this might make it easier to find the elusive mUED 
signals and try to distinguish them from supersymmetry signals 
\cite{Kundu}. Of course, the bounds on $R^{-1}$ arising from electroweak 
precision tests tell us that we cannot expect to produce KK excitations 
in the 500~GeV run. At the 1~TeV run, searches for $n = 2$ resonances 
excited by `return to the $Z$'-type effects due to ISR and beamstrahlung 
could result in narrow resonances in dilepton channels \cite{Kundu}, but 
these cannot survive the order-of-magnitude reduction in coupling 
described above. For $n = 1$ states, a detailed study made in the 
context of the 4~TeV CLIC machine \cite{Battaglia} shows that to detect 
signals with leptons and missing energy, one requires a minimum of 50 
GeV of MET, which cannot be achieved through $n = 1$ KK mode decays 
involving leptons. Prospects may be better for the signal arising from
\begin{equation}
e^+ e^- \to q_1 + \bar{q}_1 \to (q + \gamma_1) + (\bar{q} + \gamma_1)
\end{equation}
which involves two jets and MET. Here the SM backgrounds may be more 
tractable than in a hadron collider like the LHC, but a more focussed 
study is required to confirm this guess \cite{selves}.

In conclusion, then, we have shown that if a light Higgs boson is indeed 
found at the LHC, the collider phenomenology of a mUED model would be 
profoundly affected by the low cutoff demanded by the vacuum stability 
of the model. At the LHC, the $n = 1$ states will be produced in large 
numbers, but will decay into soft final states indistinguishable from 
the QCD and other SM backgrounds, except for weak trilepton and 
quadrilepton signals with accompanying jets and MET. The $n =2$ 
particles could appear as resonances in dilepton and $t\bar{t}$ final 
states, but these may be too weak to appear above the SM backgrounds. 
For $n =2$ states, a similar conclusion will hold at the ILC or CLIC 
machines, but the dijet plus MET signal has better prospects. If, 
indeed, the mUED picture of the world is correct, we may have to wait 
for the upgraded LHC to make the discovery and for the ILC to be up and 
running at 1~TeV before we can come to a definite conclusion.

{\bf Acknowledgements}: {\small The authors would like to thank the 
organisers of the WHEPP-XII (Mahabaleshwar, India), where this problem 
was conceived and discussed. AD, whose work is partially supported by 
the UGC-DRS programme acknowledges computational help from Kirtiman 
Ghosh. Thanks are also due to Dipan Sengupta (CMS Collaboration) for 
supplying some of the numbers presented in Table~\ref{tab:CShad}, to 
Avirup Shaw for helping to make some of the graphs, and to Shruti Singh 
for computer support.}


\vfill

\end{document}